\documentclass[conference, left=0.625in, right=0.625in, bottom=1in, top=0.75in]{IEEEtran}
\IEEEoverridecommandlockouts

\usepackage{cite}
\usepackage{amsmath}
\usepackage{amssymb}
\usepackage[bookmarks=false]{hyperref}
\usepackage{subfig}

\usepackage{array}
\usepackage[linesnumberedhidden, ruled,vlined]{algorithm2e} 

\usepackage[style=base]{caption}
\usepackage{optidef}
\usepackage{bm}
\usepackage{amsmath}
\makeatletter
\def\BState{\State\hskip-\ALG@thistlm}
\makeatother

\usepackage{color}
\usepackage{diagbox}
\usepackage{eso-pic}
\newcommand\AtPageUpperMyright[1]{\AtPageUpperLeft{
		\put(\LenToUnit{0.5\paperwidth},\LenToUnit{-1cm}){
			\parbox{0.5\textwidth}{\raggedleft\fontsize{10}{11}\selectfont #1}}
}}

\newcommand{\conf}[1]{
	\AddToShipoutPictureBG*{
		\AtPageUpperMyright{#1}
	}
}

\begin{document}

\title{Dynamic Channel Selection in UAVs through Constellations in the Sky}
\conf{Accepted to GLOBECOM 2019}
\author{\IEEEauthorblockN{ Guillem Reus Muns, Mithun Diddi,  Hanumant Singh,  Kaushik R. Chowdhury}
\IEEEauthorblockA{Department of Electrical and Computer Engineering, Northeastern University, Boston, USA}
\IEEEauthorblockA{Email: greusmuns@coe.neu.edu, mdiddi@coe.neu.edu, ha.singh@northeastern.edu,   krc@ece.neu.edu}}

\maketitle
\begin{abstract}
Wireless communication between an unmanned aerial vehicle (UAV) and the ground base station (BS) is susceptible to adversarial jamming. In such situations, it is important for the UAV to indicate a new channel to the BS. This paper describes a method of creating spatial codes that map the chosen channel to the motion and location of the UAVs in space, wherein the latter physically traverses the space from a given so called ``constellation point'' to another. These points create patterns in the sky, analogous to modulation constellations in classical wireless communications, and are detected at the BS through a millimeter-wave (mmWave) radar sensor. A constellation point represents a distinct n-bit field mapped to a specific channel, allowing simultaneous frequency switching at both ends without any RF transmissions. The main contributions of this paper are: (i) We conduct experimental studies to demonstrate how such constellations may be formed using COTS UAVs and mmWave sensors, given realistic sensing errors and hovering vibrations, (ii) We develop a theoretical framework that maps a desired constellation design to error and band switching time, considering again practical UAV movement limitations, and (iii) We experimentally demonstrate jamming resilient communications and validate system goodput for links formed by UAV-mounted software defined radios.
\end{abstract}


\section{Introduction}
Unmanned aerial vehicles (UAVs) are utilized for military operations, surveillance, disaster management, telecommunications, monitoring, and cargo delivery  \cite{Mozaffari2018}. All such roles require continuous control, navigation, communication and autonomy \cite{Wang2017}, requiring robust links between the UAV itself and the ground base station (BS) \cite{Mozaffari2016, Seo2017} or between distributed UAVs \cite{Trotta2018}. The degradation in wireless links caused by adversarial actions, such as jamming or interference, has been widely studied for diverse applications, such as WSN \cite{Xu2005} or wireless charging \cite{Naderi2014}.

\begin{figure}[t!]
  \centering
   \includegraphics[width=0.4\textwidth]{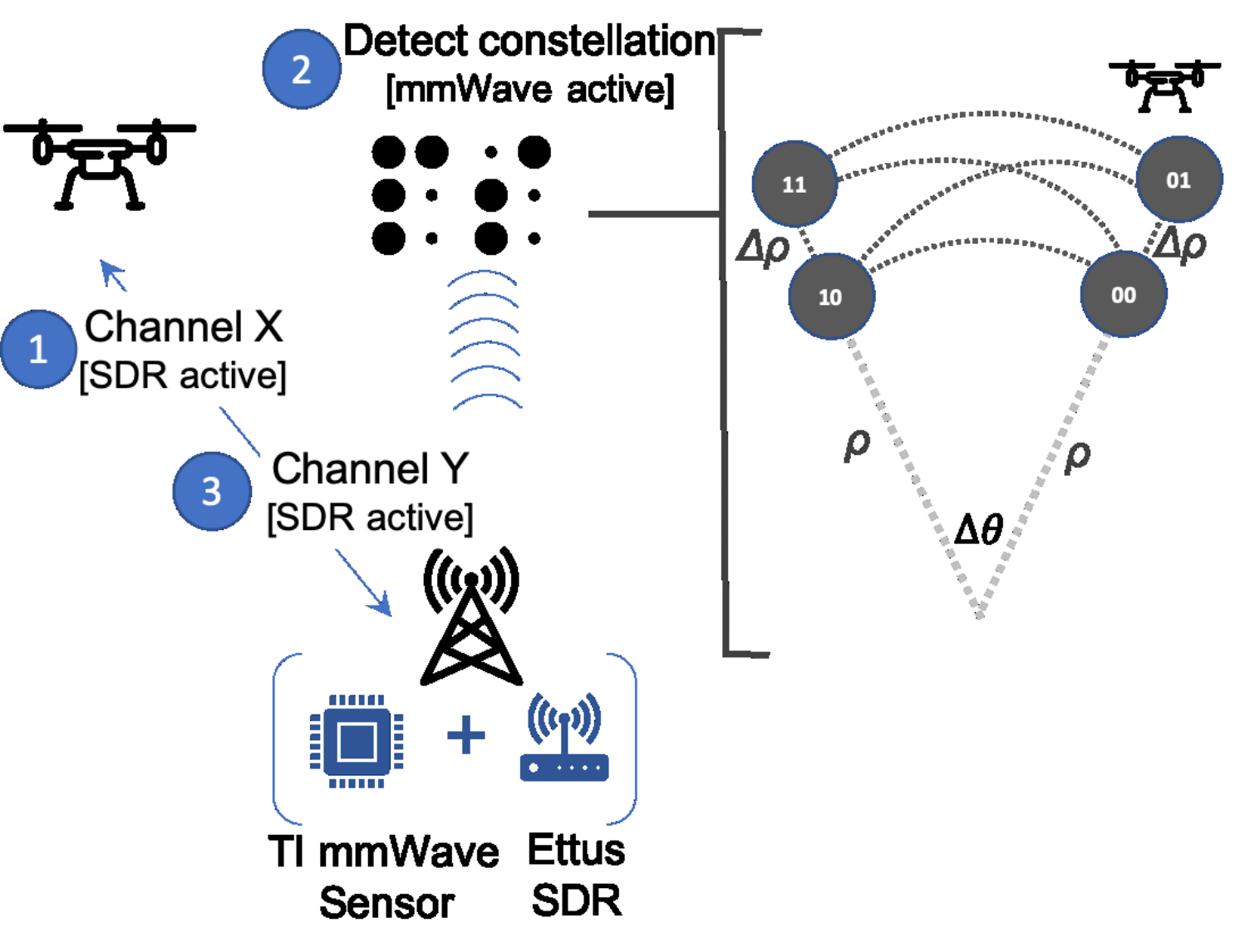}
  \caption{UAV communicates with ground station on Channel X (step 1). When jammed, it moves through physical space to encode new channel information in a constellation, detected by mmWave sensor (step 2), expanded for 4-physical locations. The communication link switches to Channel Y free from jammer (step 3).} 
  \label{fig:diagram}
  \vspace{-15pt}
\end{figure}

\noindent$\bullet$\textbf{Overview of the approach:} Fig. \ref{fig:diagram} shows a sample scenario where the RF link on channel X between the UAV and the BS is severed due to a jamming attack. The UAV selects a new channel Y for continuing the communication, but is unable to let the BS know of its choice owing to the active jammer. So, it uses an out-of-band control signaling method involving relaying channel information by moving between different spatial locations. Notice that X and Y could represent any available transmission band that the hardware of both the UAV and the BS could support (i.e. sub-6GHz, mmWave, etc), making this solution applicable to broadband jammers working in a certain band. Information conveying modulation constellations are used in classical wireless communications, and our approach attempts to map a similar concept into the UAV scenario. This creates a low-bandwidth control channel that is resilient to the ongoing jamming attack. Note that this approach would remain secure against a jammer equipped with its own localization technology since the location-channel mapping would be unknown on its side. Our approach relies on accurate localization of the UAV in 2-D space (in fact, any imprecision results in symbol error at the BS). While  sensing-aided communications system have been explored in other works \cite{Reusmuns2019}, for this paper we choose a single-chip Frequency-Modulated Continuous Wave (FMCW) mmWave radar.

\noindent$\bullet$\textbf{Research challenges:} The idea of using spatial constellations raises many unique research challenges at the intersection of wireless communication and robotics. Firstly, based on an experimental study using a COTS mmWave radar TI IWR1642, we identify the regions where the sensor accuracy drops. This results is generating non-intuitive and irregular shapes for the resulting constellation. For instance, in Fig. \ref{fig:diagram}, a regular QPSK modulation used in  classical RF would have its points at the four vertices of a square, whereas our approach traces arcs in the sky for the same points. We answer the fundamental question of how these physical constellations scale and what forms they take as the number of bits required to represent additional information also changes.

With the available degrees of spatial freedom, we must also determine the separation between points, defined by $\Delta_\rho$-$\Delta\theta$, which represent the symbol spacing between any consecutive constellation points along the $\rho$ or $\theta$ polar coordinates axis, respectively. The need of using polar coordinates is explained later in this paper. Moreover, the problem of inter-point spacing has many non-intuitive elements. Since the UAV must physically move from one point to another, the separation between the constellation points may be minimized to reduce the travel time, and thus increase the information capacity. This is a distinction not present in classic information constellation designs, where the separation between symbols is always maximized to reduce the BER. However, simply bunching the points very close causes two problems: The natural hovering and instability during flight can move the UAV close to an incorrect location. It also decreases the ability of the ground-based mmWave sensor to resolve the UAV locations at these discrete points. Overall, designing such a physical constellation based control signaling method involves many unique interdisciplinary conditions at the intersection of robotics, communication and sensing.

In summary, the main contributions in this paper are:
\begin{enumerate}
    \item We introduce the concept of spatial modulation constellation for UAVs and motivate its application as a method for frequency band selection for jamming resilience.
    \item Through experimental traces and characterization of the mmWave sensor, we design a two-step clustering algorithm that is able to process the positional data and distinctly identify different UAVs with minimal impact of noise.
    \item We design a constellation scheme for N=2 points in space, and propose generalization steps, by taking into account mmWave sensor performance and UAV flight limitations. The approach identifies the optimal separation distance that requires minimum movement for the UAV, while ensuring robustness in detection.
    \item We experimentally demonstrate the jamming resilience and implement our design on DJI M600 UAVs with Ettus B210 software defined radios for 2 constellation points. Additional simulation results are provided for larger constellation sizes to demonstrate scalability.
\end{enumerate}
\vspace*{-3pt}

\section{mmWave Sensing for UAV Localization}
We use a Texas Instruments IWR1642 evaluation module that has 10.4x10.4mm mmWave sensor incorporating FMCW radar technology. The sensor works in the 76-81GHz band with up to 4GHz chirp, and feeds real time location information to a laptop that analyzes the resulting point cloud. Also, we use a DJI Matrice M600 Pro UAV with access to the low level flight controller telemetry data. Furthermore, we integrated a real-time GPS kinematic solution, called DJI D-RTK, in the UAV for cm-level GPS accuracy, compared to variations in the scale of $\pm 1.5$m in the horizontal plane otherwise.

\begin{figure}[t!]
\centering
	\subfloat[]{%
  	\includegraphics[width=0.49\linewidth]{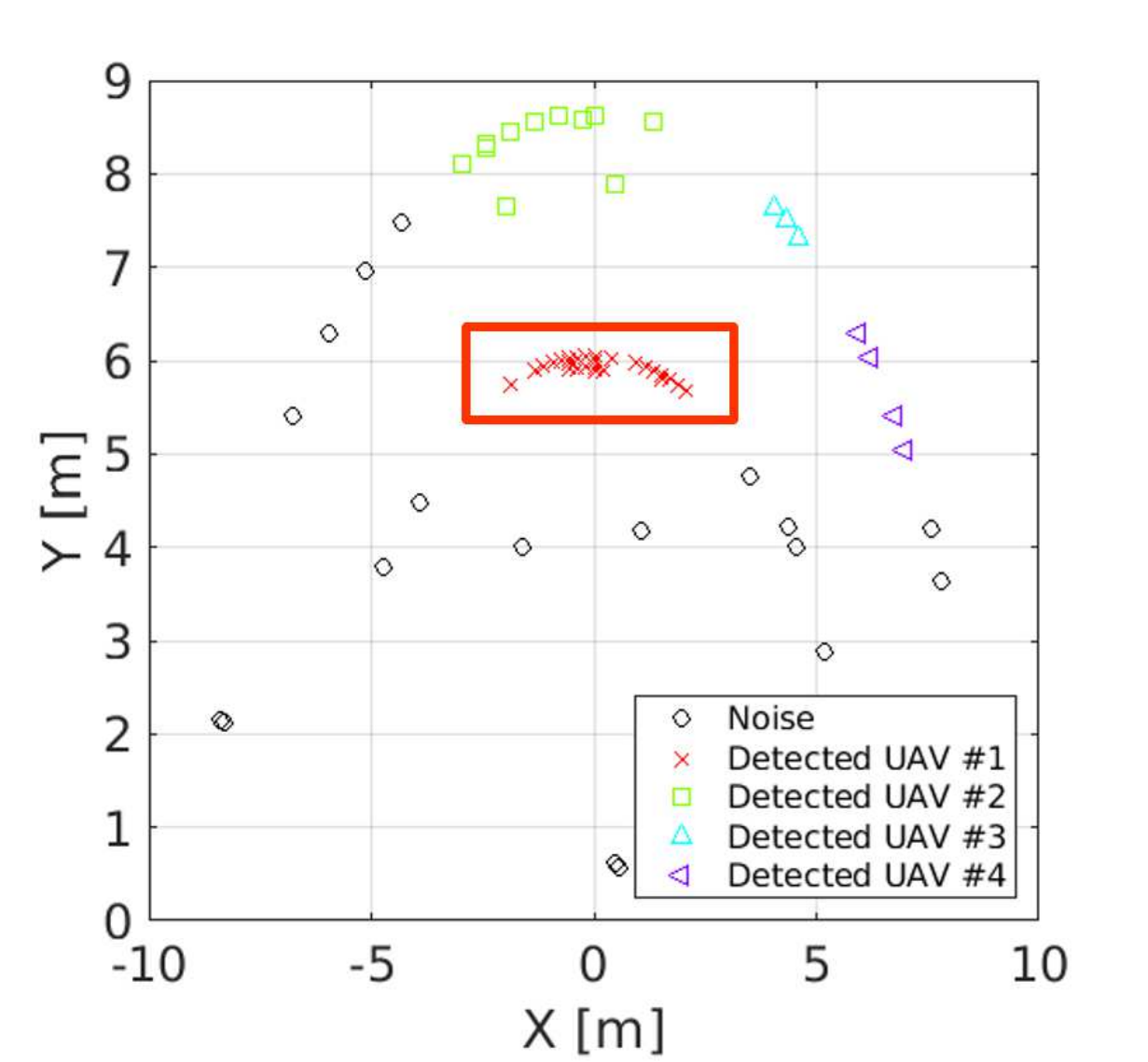}\!\!\!\!
  	\label{subfig:cls1uavf}%
	}
	\subfloat[]{%
  	\includegraphics[width=0.49\linewidth]{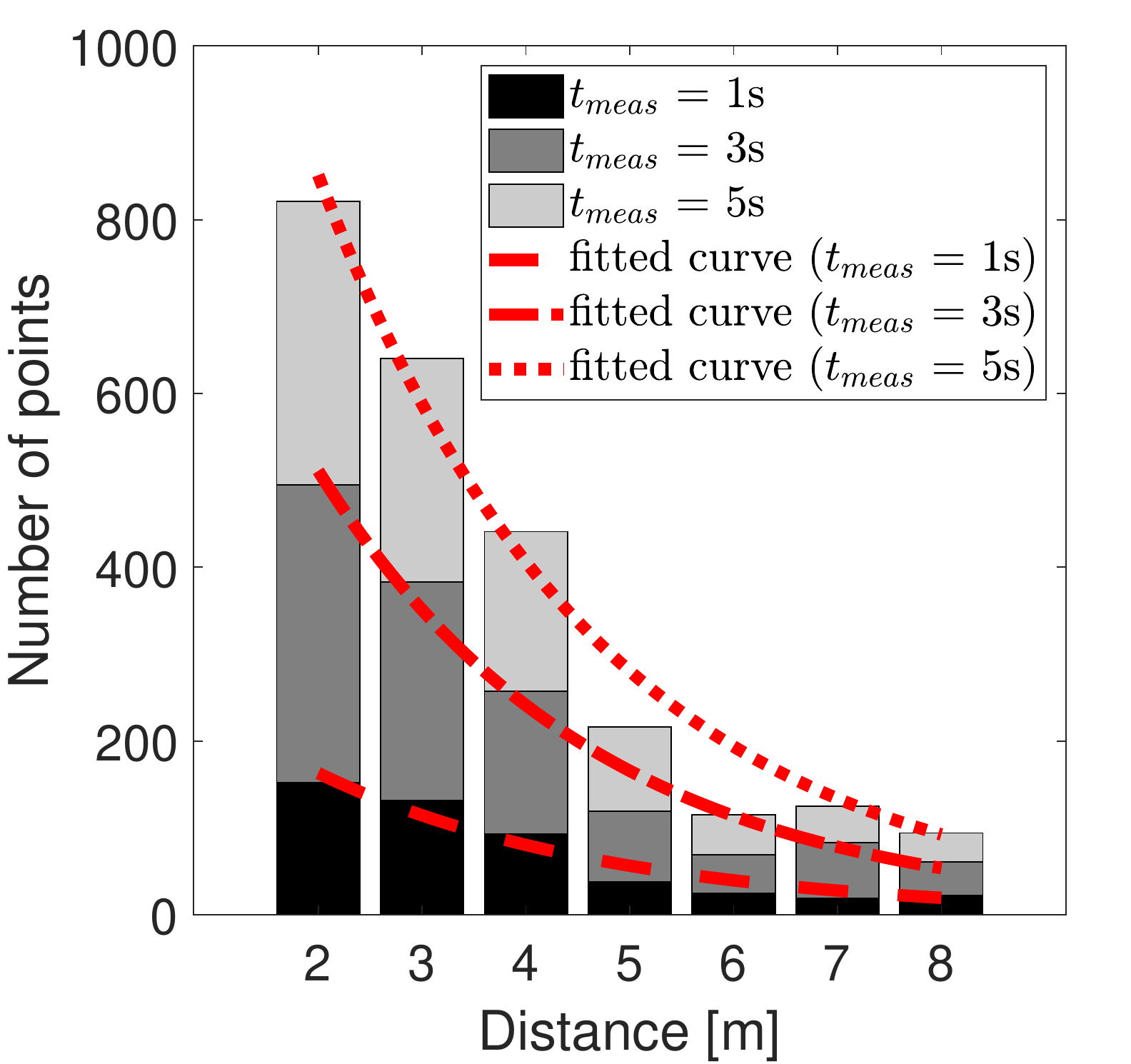}%
  	\label{subfig:FC}%
	} 
\caption{(a) Measured point cloud with a UAV at position [0,6]m shows considerable noise and misclassification. However, after setting \textit{MinPts} based on our analysis in Sec.~\ref{subsec:Hi-DBSCAN}, we successfully obtain a cluster  of feasible points around the UAV's location (enclosed in the red bounding box).}
\vspace{-20pt}
\label{fig:staticuav}
\end{figure}

\subsection{Static UAVs}
\label{subsubsec:arcedcl}
\begin{figure}[t!]
\centering
	\subfloat[]{%
  	\includegraphics[width=0.49\linewidth]{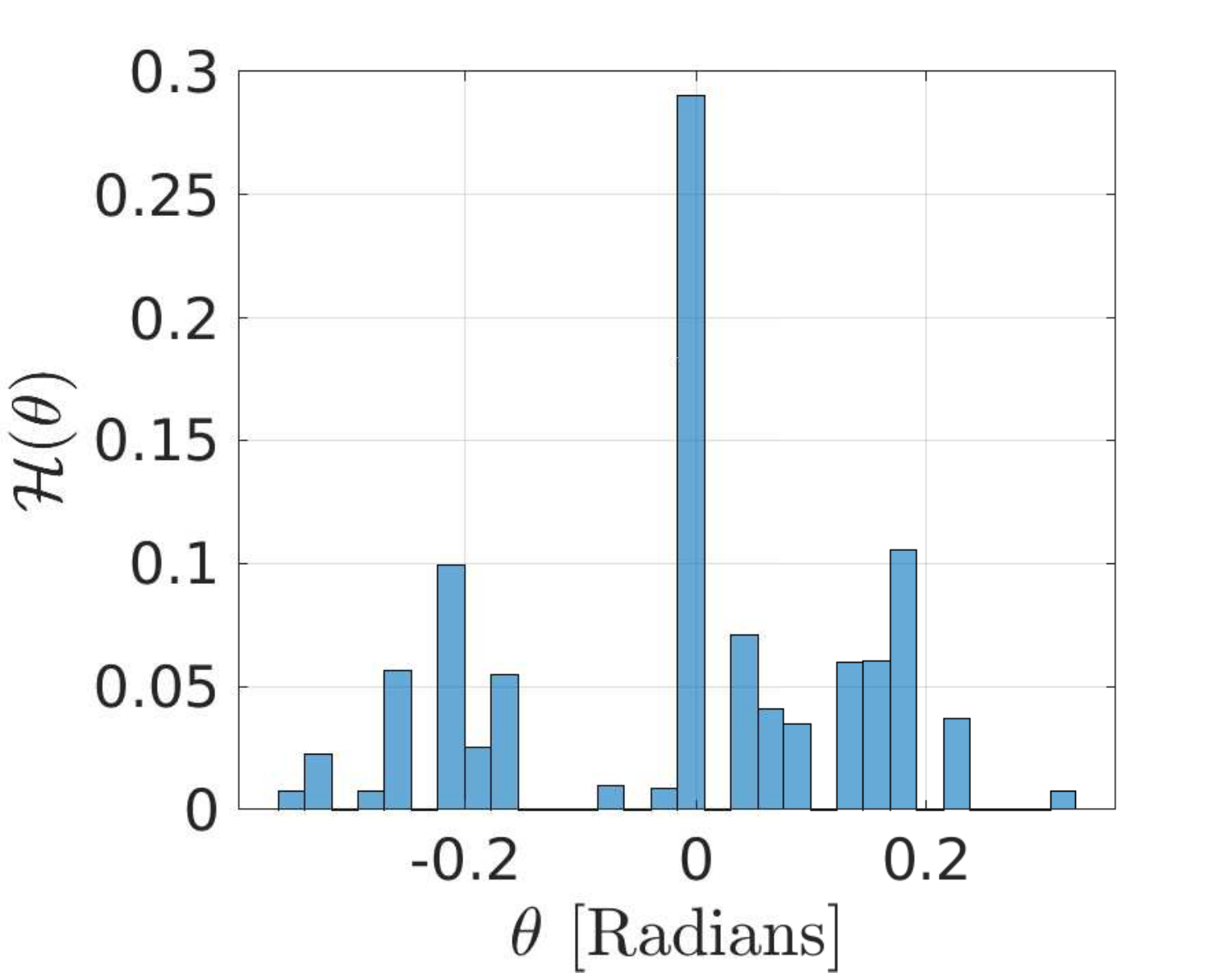}\!\!\!\!\!\!\!\!
  	\label{subfig:hTh}%
	}
	\subfloat[]{%
  	\includegraphics[width=0.49\linewidth]{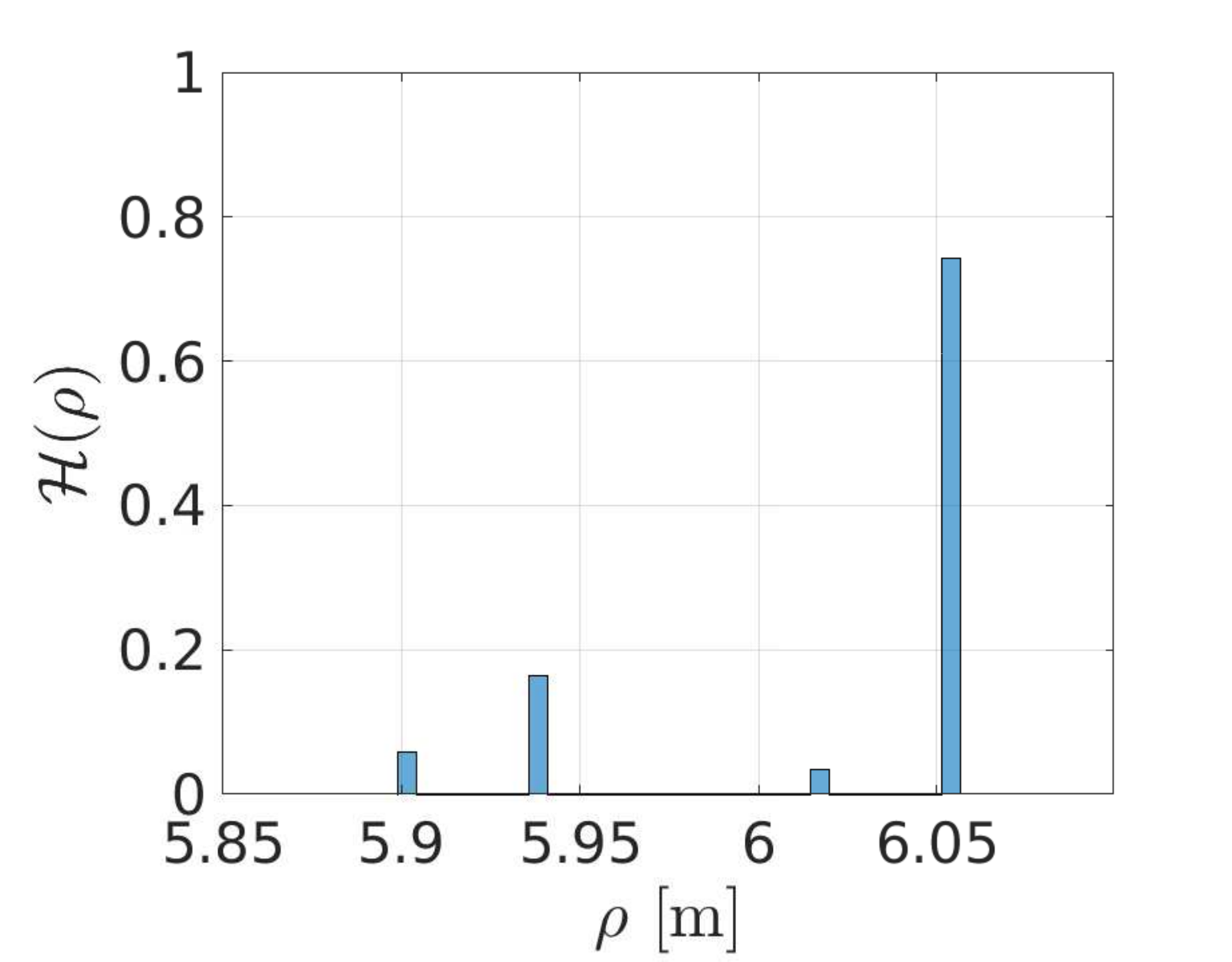}%
  	\label{subfig:hRho}%
	}
\caption{Histogram for $\theta$ (a) and $\rho$ (b) for the UAV point cloud in Fig.~\ref{subfig:cls1uavf}. Both variables exhibit a peak which is leveraged for refining position estimation.}
\label{fig:hists}
\vspace{-15pt}
\end{figure}

Consider a UAV statically supported by a tripod, approximately 1m from the ground, and placed at the coordinate [0,6]m with respect to the origin [0,0] in the x-y plane, where the sensor is located. The UAV propellers are set to rotate at low rpm, and the sensor is configured to only detect moving objects. The sensor reports valid spatial coordinates (see red point cloud in Fig.~\ref{subfig:cls1uavf}), but also many additional noise readings. Interestingly, the point cloud is not uniformly distributed around the target. Using polar coordinates, we see the sensor is more accurate in terms of distance from origin (say, $\rho$) rather than angle of the target \textit{wrt} to origin (say, $\theta$). Indeed, the histogram of the point cloud shown in Fig.~\ref{fig:hists} validates the comparatively greater uncertainty in localization accuracy with respect to $\theta$ over $\rho$. This key insight is used for spacing the constellation points in our approach, which results in an asymmetric form of the constellation.

\vspace{-2pt}
\subsection{Hovering UAVs}
\label{subsec:pdfs}
\begin{figure*}[h!]
\centering
	\subfloat[]{%
  	\includegraphics[width=0.26\linewidth]{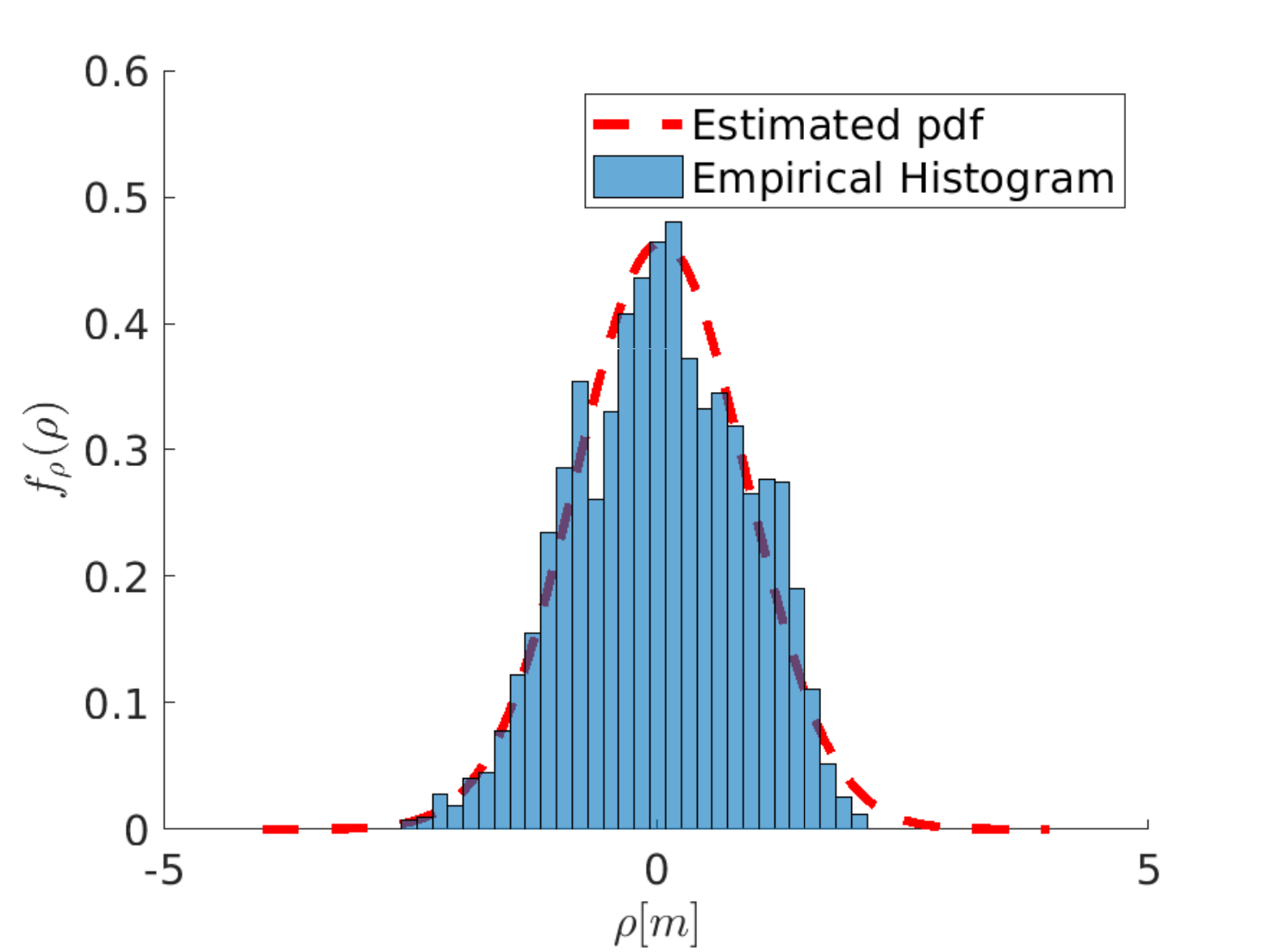}\!\!\!\!
  	\label{subfig:pdfRho}%
	}
	\subfloat[]{%
  	\includegraphics[width=0.25\linewidth]{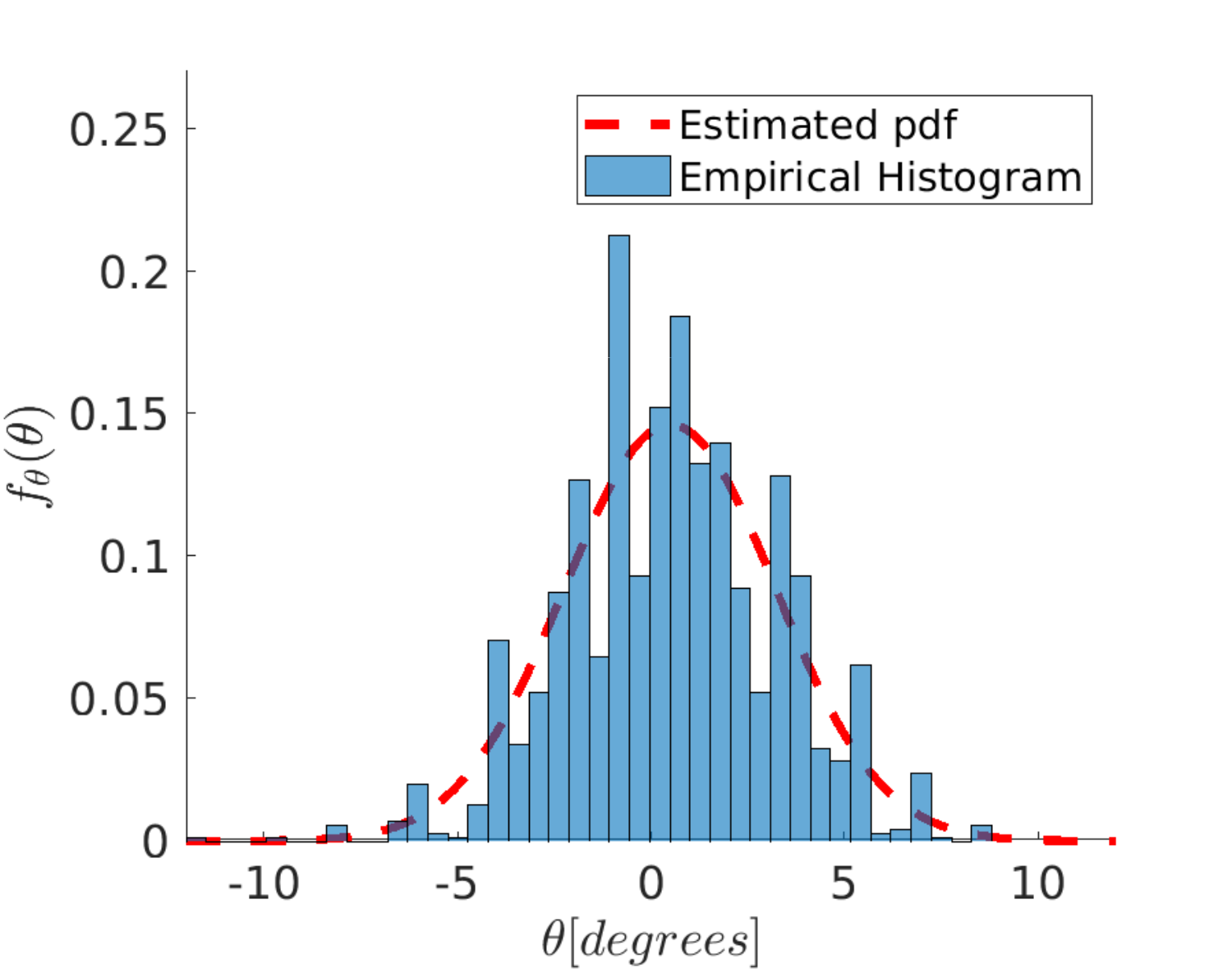}\!\!\!\!
  	\label{subfig:pdfTh}
	}
	\subfloat[]{%
  	\includegraphics[width=0.26\linewidth]{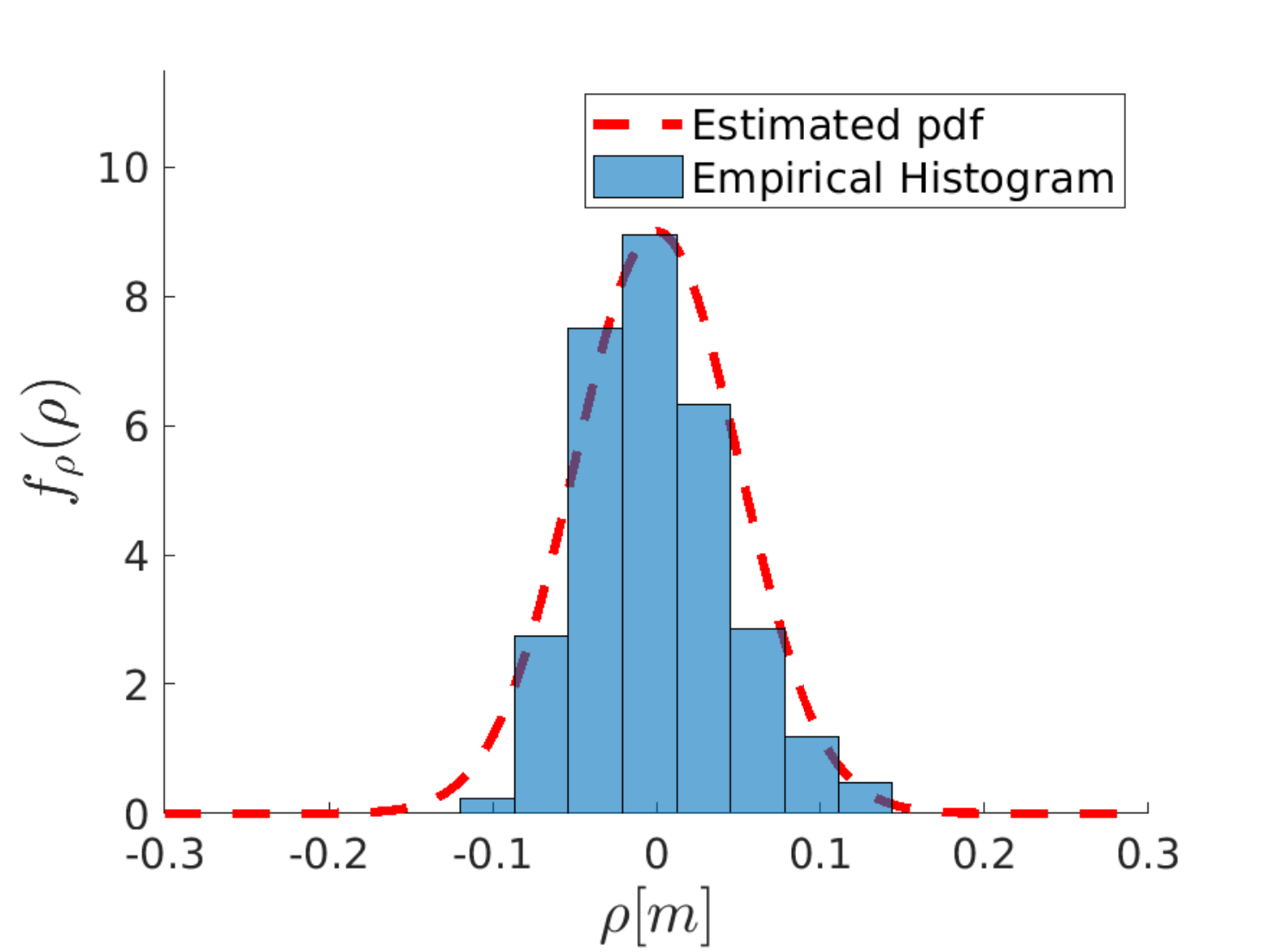}\!\!\!\!
  	\label{subfig:pdfRhoRTK}%
	}
	\subfloat[]{%
  	\includegraphics[width=0.24\linewidth]{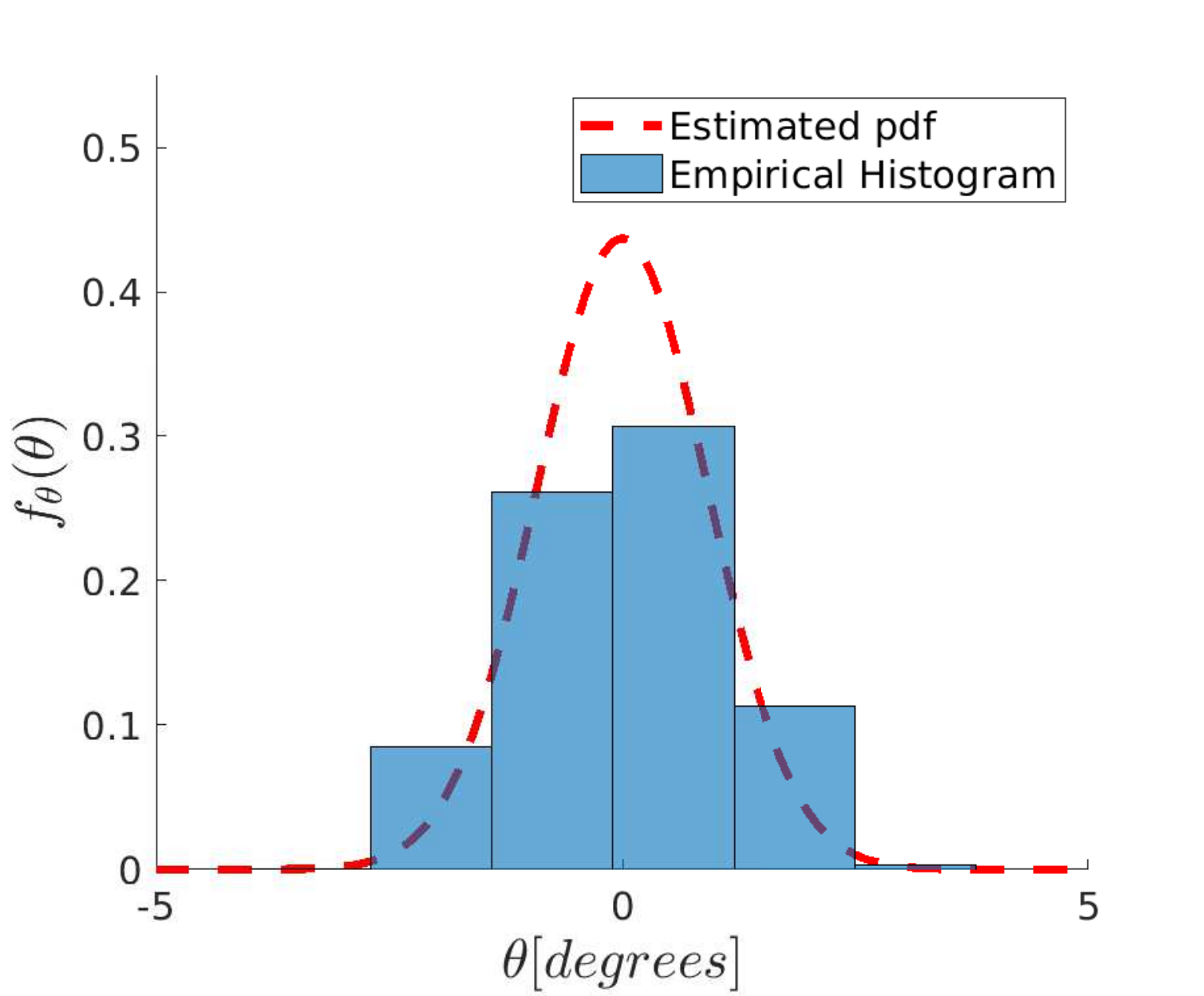}%
  	\label{subfig:pdfThRTK}%
	}
\caption{Probability density function for $\rho$ (a) and $\theta$ (b) using standard GPS localization. Probability density function for $\rho$ (c) and $\theta$ (d) using RTK localization.}
\label{fig:pdfs}
\end{figure*}

When a UAV is set to operate a given point in 3-D space, it shows slight displacement over time in all three dimensions. We next determine if this unpredictable hovering motion can potentially result in the mmWave sensor mis-detecting the target constellation point. For the purpose of this work, we focus on a 2-D plane. We collect measurements while flying the UAV at coordinates with different $\rho$ values (fixing $\theta$ at $0^{\circ}$) from 2m to 12m in steps of 0.5m with staying duration of 2 minutes per point. The same experiment is repeated for different $\theta$ coordinates, from $-32^{\circ}$ to $32^{\circ}$ in steps of $4^{\circ}$, keeping $\rho$ constant. The benefits of using D-RTK can be immediately seen in Fig.~\ref{fig:pdfs}, giving an increased accuracy of both $\rho$ and $\theta$ estimation, respectively, when the RTK is active in (c) and (d), versus using classical GPS in (a) and (b). Additionally, we leverage the fact that the distribution of the error along both variables (\textbf{$\rho$} and \textbf{$\theta$}) follow a Gaussian distribution in Sec.~\ref{sec:pe}.

\vspace{-2pt}
\subsection{Accurate Real-time Localization}\label{subsec:Hi-DBSCAN}

Our goal here is to accurately identify the constellation point from the data clouds obtained by the mmWave sensor. We use DBSCAN (Density Based Spatial Clustering of Applications With Noise) \cite{Ester1996} as the starting point. DBSCAN groups together sets of points based on the region density, and at the same time, is able to detect outliers with low run-time overhead. 
In addition, it can be used in a wide range of cluster shapes (i.e. linear, concave, circular, etc.).  
We make two main contributions here: (i) we accurately initialize the parameters of the DBSCAN algorithm using measurement studies which are also seen in this section, and (ii) we reduce uncertainty with a novel \textit{weighted histogram} approach to create Hi-DBSCAN, which increases accuracy over the stock algorithm. 

\noindent$\bullet$\textbf{DBSCAN tuning for UAVs:} We focus on three main parameters for the proposed  Hi-DBSCAN: (i) $\epsilon$, a measure of radius that defines the circular neighborhood around the true center of the UAV. Any measurement point within this circle is called as an $\epsilon$-neighbor. (ii) \texttt{MinPts}, the minimum number of neighboring points a true UAV location should have in order to not be classified as noise. (iii) \texttt{Dist}, the maximum distance at which the mmWave sensor should detect measurement data. $\epsilon$ and \texttt{MinPts} are legacy DBSCAN parameters which would result in undesirable performance if not properly tunned \cite{Schubert2017}. $\epsilon$ can be trivially set from the dimensions of the UAV.  Similarly, we can set the UAV flight boundaries to directly compute \texttt{Dist}. \texttt{MinPts}, on the other hand, is a function of the frame rate of the sensor (i.e., samples produced per second), the time over which the samples are collected $t_{meas}$), and the total number of points obtained \texttt{R}, which include legitimate signal reflections from the UAV, along with noise and radar artifacts.

For a fixed frame rate, \textit{R} increases linearly over time, as expected. Moreover, as Fig.\ref{subfig:FC} shows, \textit{R} is also distance dependent. In order to  characterize its dependency, we fit an exponential function curve to our data ($\mathcal{F}=a e^{bd}$, where \textit{d} is distance and \textit{a} and \textit{b} are the parameters to be estimated). Fig.~\ref{subfig:FC} shows the obtained fitting curves for different $t_{meas}$ values. Thus, UAVs at different distances create different density point clouds. The lowest  density areas give the lower bound on the performance of the clustering algorithm (i.e., if the UAV is detected accurately in low density areas, it is highly probable that will also be detected in more dense areas.). Thus, we define  \texttt{MinPts} as $ \mathrm{ \texttt{MinPts}} = \alpha \mathcal{F}(t_{meas}, \mathrm{\texttt{Dist}})$,
 where $\alpha$ is a scaling factor. In our experiments. We set $\alpha$ to 0.5 under the assumption that at least half of the detected points are contributed by the UAV at a given location. Fig.~\ref{subfig:cls1uavf} shows how properly tuning the \texttt{MinPts} parameter directly identifies the candidate points (in the red box) while \textit{all} other extraneous points are labelled as noise, unlike the case with a naive \texttt{MinPts} value, where the clustering predicts there are 3 extra UAVs.

\noindent$\bullet$\textbf{Weighted histogram analysis:} DBSCAN outputs clusters, each containing a cloud of candidate UAV location points. However, our goal is to ultimately estimate a single location for the UAV. Thus, we add a second stage, which aims to refine the position estimation by performing a weighted histogram analysis. For every point cloud we make the following observation:  (i) Points with higher received power are more likely to be the true location of the UAV, and (ii)  Points in every cluster are more densely distributed around the true location. Thus,  we define a weighted histogram, using the received power as the weights: 

\begin{equation}
        \mathcal{H}_i = \frac{n_i}{\sum\limits_{n=1..N}  p_n\times w_n}
\end{equation}

\begin{equation}
        n_i = \sum_{i\in I} p_i\times w_i 
\end{equation}

Where $\mathcal{H}_i$ is the weighted histogram, $n_i$ is the sum of the weighted points at bin \textit{i}, \textit{N} is the number of points in the cluster, \textit{I} defines the range of bin \textit{i} and $w_i$ is the weight applied to every point $p_i$. 
Consequently, the final estimated point is $\mathbf{d} = [\theta_e , \rho_e]$, where  $\theta_e = \max_\theta \mathcal{H}(\theta)$ and $\rho_e = \max_\rho \mathcal{H}(\rho)$. In Fig.~\ref{fig:hists} we see how both $\mathcal{H}(\rho)$ and $\mathcal{H}(\theta)$ show a clear peak, which is known to match the UAV true location.

\section{Constellation Location Error Probability}
\label{sec:pe}
The unpredictable hovering introduces errors in the localization process. We model the location of each UAV as a random variable defined as $\textbf{p}= (\theta, \rho)$ using the polar coordinate system. As discussed in Sec.~\ref{subsec:pdfs}, the UAV movement along each axis follows a Gaussian distribution. Assuming statistical independence, we discuss the means and variances of these variables next: Let $\textbf{s}_i = (s_{\theta_i}, s_{\rho_i})$ be the polar coordinates for a given symbol \textit{i} in a 2D plane. 
In an ideal case, the UAV location exactly overlaps with a given symbol location, or at least, exhibits a mean location $\boldsymbol{\mu}_i $ equal to the constellation point, i.e.,  $\boldsymbol{\mu}_i = \textbf{s}_i$.  Furthermore, let $\boldsymbol{\sigma}_i = (\sigma_{\theta_i}, \sigma_{\rho_i})$ denote the vector of standard deviations that defines the precision with which a given UAV hovers around certain symbol coordinates. The lower the $\boldsymbol{\sigma}_i$, the more stable the UAV is. Since the UAV's position $\textbf{p}_i $ on any axis is assumed to be an independent Gaussian random variable, while transmitting symbol \textit{i}, this  can be expressed as:
\begin{align}
\textbf{p}_i &= \mathcal{N}(\boldsymbol{s}_i, \boldsymbol{\sigma}_i\mathcal{I} ),\qquad \forall i \in \{1,2,...,N\}
\end{align}
where \textit{N} stands for the number of different symbols in the constellation. There is a symbol error anytime the hovering displaces the UAV out of its feasible symbol region. Without loss of generality, each symbol region $\mathcal{R}_i$ is defined as:
\begin{equation}
\mathcal{R}_i = \begin{cases}
				\alpha_i^l < \theta < \alpha_i^u \\
                \beta_i^l < \rho < \beta_i^u 
				\end{cases}
\end{equation}
where $\alpha$ and $\beta$  define the bounds on each of the axis and the \textit{l} and \textit{u} respectively stand for the \textit{lower} and \textit{upper} bound that define the $\mathcal{R}_i$ limits. While we provide a detailed analysis for the case of N=\{2\} next, higher order constellation sizes are not included due to space constraints, though they follow very similar steps. As the constellation design is not only dependent on \textit{N} but $\Delta_{\rho}$ and $\Delta_{\theta}$, a general symbol error probability is hard to derive. However, in Sec.~\ref{subsec:PeGenForm} we develop a generic framework that accounts for the number of neighbors of each symbol in order to compute the average probability of error.

\subsection{Analysis for N=2}
\label{subsec:N2}
The constellation can have two possible configurations, where both symbols are placed either along the $\rho$ or the $\theta$ axis. In order to avoid redundancy, we derive the expression for the $\theta$ case only, where $s_\theta=\{-\frac{\Delta_\theta}{2}, \frac{\Delta_\theta}{2}\}$. Here, we define the probability of error as:

\begin{align}
    P_e &= \sum_{i=\{1,2\}}P(\textbf{p}_i \notin \mathcal{R}_i )= \sum_{i=\{1,2\}} P(\beta_i^l > \theta_i > \beta_i^u) P(s_i)\\
    &= \sum_{i=\{1,2\}} P(\beta_i^l > s_{\theta_i} + \varepsilon_{\theta_i} > \beta_i^u) P(s_i)
\end{align}

We reformulate $\textbf{p}_i$ as the addition of a deterministic value ($\textbf{s}_i$) and a random component ($\boldsymbol{\varepsilon} = \mathcal{N}(0, \boldsymbol{\sigma}_i\mathcal{I})$). As the symbols are equally probable with same error probabilities:

\begin{align}
    P_e = \frac{1}{2} 2  P(\varepsilon_{\theta_i} > \beta_i^u-s_{\theta_i})  = \mathcal{Q}\left(\frac{\beta_1^l-s_{\theta_1}}{\sigma_{\theta_1}}\right)
\end{align}

Where $\mathcal{Q}(x) = \int_x^\infty \frac{1}{\sqrt{2\pi}}\exp{\frac{-t^2}{2}}$ is adopted to simplify the expression. Finally, considering $s$ as the constellation point centered at $\theta=0$, then $\beta_1^l=0$, $P_e = \mathcal{Q}\left(\frac{\Delta_\theta/2}{\sigma_{\theta_1}}\right)$.


Similarly, if both symbols were placed along the $\rho$ axis $P_e = \mathcal{Q}\left(\frac{\Delta_\rho/2}{\sigma_{\rho_1}}\right)$.
This result matches with the BPSK probability of error through an AWGN channel, as expected.

\subsection{Generalized formulation}
\label{subsec:PeGenForm}

We derived the error probability for the constellation size (N=2) in Sec.~\ref{subsec:N2}. We next tabulate the error probability as a function of the number of neighbors a given constellation point has along $\rho$ axis ($n_\rho$) and $\theta$ axis ($n_\theta$).  Also, a given symbol is considered as the neighbor of another one if it is placed at a distance $\Delta_{\rho}$ - $\Delta_{\theta}$ along the $\rho$ - $\theta$ axis respectively. We define the probability of error in terms of $n_\rho$ and $n_\theta$ because this allows us to derive and expression for every single symbol in the constellation and then compute total $P_e$ as the average. Table.~\ref{tab:Pe} shows $P_e(n_\theta,n_\rho)$ for every possible $(n_\theta,n_\rho)$ pair.

\begin{table}[h!]
    \centering
    \begin{tabular}{|c|c|c|}
         \hline
         \textbf{$n_\theta$} & \textbf{$n_\rho$} & \textbf{$P_e(n_\theta,n_\rho)$}\\
         \hline
         \hline
         2&2&$1 - \left\{ \left[1 - 2\mathcal{Q}\left(\frac{\Delta_\theta/2}{\sigma_{\theta_1}}\right) \right]  \right. \left. \left[1- 2\mathcal{Q}\left(\frac{\Delta_\rho/2}{\sigma_{\rho_1}}\right) \right] \right\}$\\
         2&1&$1 - \left\{ \left[1 - 2\mathcal{Q}\left(\frac{\Delta_\theta/2}{\sigma_{\theta_1}}\right) \right]  \right. \left. \left[1- \mathcal{Q}\left(\frac{\Delta_\rho/2}{\sigma_{\rho_1}}\right) \right] \right\}$\\
         2&0&$2\mathcal{Q}\left(\frac{\Delta_\theta/2}{\sigma_{\theta_1}}\right)$\\
         1&2&$1 - \left\{ \left[1 - \mathcal{Q}\left(\frac{\Delta_\theta/2}{\sigma_{\rho_1}}\right) \right]  \right. \left. \left[1- 2\mathcal{Q}\left(\frac{\Delta_\rho/2}{\sigma_{\theta_1}}\right) \right] \right\}$\\
         1&1&$ 1 - \left\{ \left[1 - \mathcal{Q}\left(\frac{\Delta_{\theta}/2}{\sigma_{\theta}}\right) \right]  \right. \left. \left[1- \mathcal{Q}\left(\frac{\Delta_{\rho}/2}{\sigma_{\rho}}\right) \right] \right\}$\\
         1&0&$\mathcal{Q}\left(\frac{\Delta_\theta/2}{\sigma_{\theta_1}}\right)$\\
         0&2& $2\mathcal{Q}\left(\frac{\Delta_\rho/2}{\sigma_{\rho_1}}\right)$ \\
         0&1&$\mathcal{Q}\left(\frac{\Delta_\rho/2}{\sigma_{\rho_1}}\right)$\\
         \hline
    \end{tabular}
    \vspace{5pt}
    \caption{$P_e$ expressions for different $(n_\theta,n_\rho)$ pairs.}
    \label{tab:Pe}
\end{table}

Given a constellation setup and the probability of error for every symbol, the total probability of error for any constellation can be expressed as:

\begin{equation}
    P_e = \frac{1}{N} \sum_{n=1}^N P_e^n(n_\theta,n_\rho)
    \label{eq:PeGeneral}
\end{equation}

where $P_e^n$ represents the $P_e$ for symbol \textit{n}. Notice how every expression in Table.~\ref{tab:Pe} depends explicitly on $\Delta_\rho$ and $\Delta_\theta$. This will be relevant in the following section, where the constellation design is explained.

\section{Creating Efficient Constellations}
\label{sec:const}

In this section, we describe in detail how to design the constellations composed of the set of $N$ points given by $\text{\textbf{p}}^{*}$ (see Fig.~\ref{fig:optimal2}). We aim to minimize the average travel time ($\mathcal{T}$), which occurs when the inter-symbol distance is also minimized. 
The parameters $\Delta_\rho$ and $\Delta_\theta$ are selected from the analysis in Sec.~\ref{sec:pe} using the generalization of error probability obtained in Eq.~\ref{eq:PeGeneral} and Table.~\ref{tab:Pe}. As mentioned above, the error directly depends $\Delta_\rho$ and $\Delta_\theta$. Then, both of these parameters are set so that $P_e$ is kept below a certain threshold $\xi$. Thus, the problem can be formulated as:

\vspace{-3pt}
\begin{subequations}
\begin{align}
\underset{\bf{p}^{*}}{\text{min}} & \quad \mathcal{T}(N,\Delta_\rho, \Delta_\theta)\label{eq:obj} \\ 
\text{such that: } & P_e \leq\xi
\end{align}
\vspace{-5pt}
\end{subequations}
\vspace{-3pt}

where $\bf{p}^{*}$ are the \textit{N} optimal points with spacing $\Delta_\rho$ and $\Delta_\theta$ that minimize $\mathcal{T}$ while ensuring $P_e \leq \xi$ are obtained.

\subsection{Exhaustive exploration}

\begin{figure}[t!]
\centering
  	\includegraphics[width=0.8\linewidth]{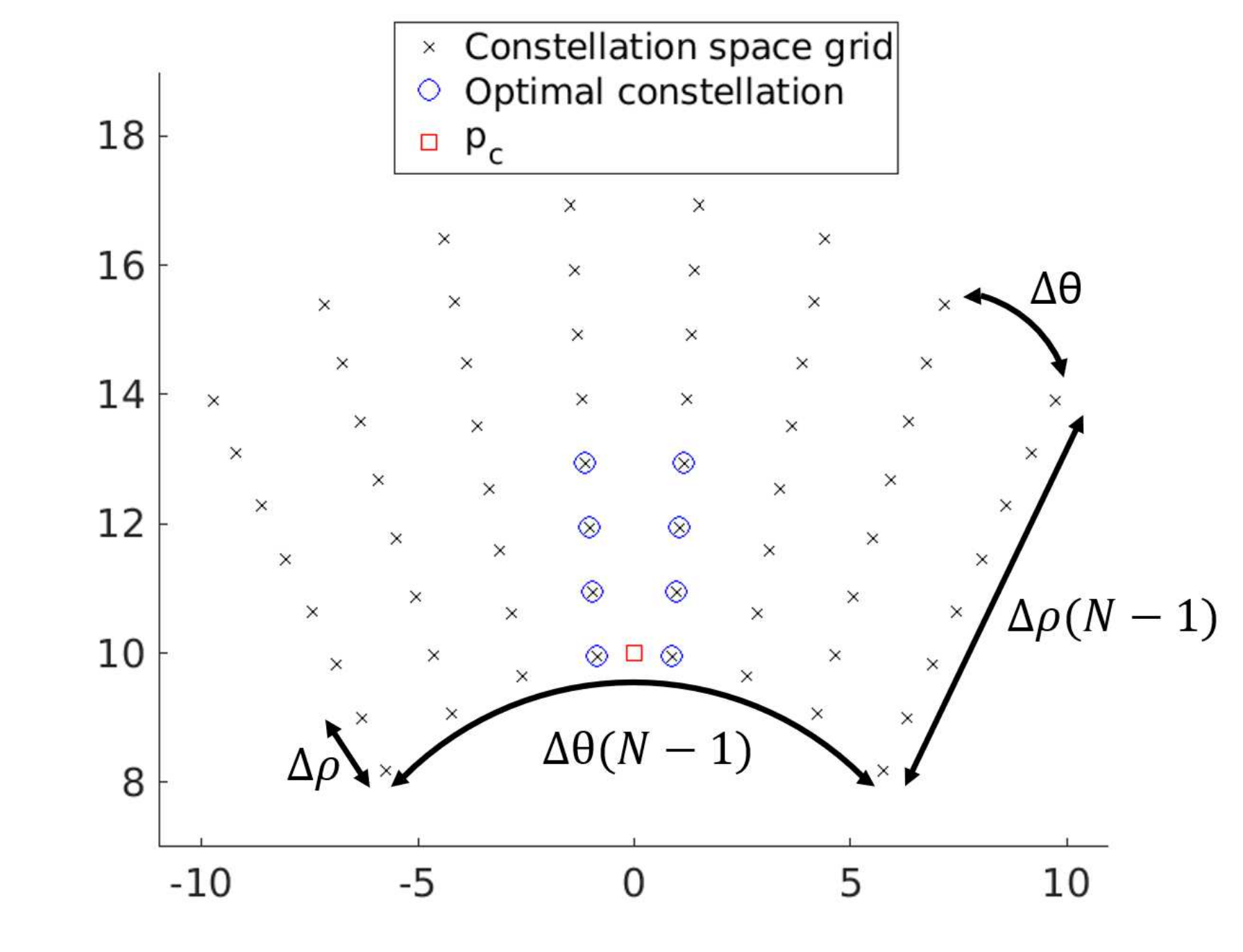}
\caption{Optimal constellation}
\label{fig:optimal2}
\vspace{-10pt}
\end{figure}

From Sec.~\ref{subsubsec:arcedcl}, the mmWave sensing accuracy is not equally distributed along both the polar coordinate axes. For instance, considering Fig.~\ref{fig:optimal2}, the optimal constellation for N=8, uses 4 values in the $\rho$ axis while only 2 along $\theta$ axis. If we analyze this asymptotically, the optimal design may require placing all the \textit{N} points along one of the axis. Thus, we define a grid of L=$N^2$ elements which contains all the possible solutions for our problem. For a given starting point ($\boldsymbol{p_c}=[\theta_c, \rho_c]$), the grid range is defined as $\{[\theta_c -\frac{\Delta_\theta(N-1)}{2}, \theta_c +\frac{\Delta_\theta(N-1)}{2}], [\rho_c, \rho_c +\Delta_\rho(N-1)]\}$ (see Fig.~\ref{fig:optimal2}).
The optimal constellation ($\bf{p}^{*}$) is the set of \textit{N} elements out of \textit{L} for which its average distance is minimized, i.e., from (\ref{eq:obj}). To solve this, we propose (i) an exhaustive search for pre-flight computationally non-restricted scenarios, and (ii) a heuristic algorithm for in-flight constellation calculation. Moreover, we show how our heuristics converge to a close to optimal solution and analyze the complexities.

\subsection{Heuristic exploration}
In order to reduce the complexity of an exhaustive search that compares the average distance for any set of $N$ points in a grid with $L$ elements, we assume that the inter-symbol distance is reduced if the set of points closer to $\boldsymbol{p_c}$ are picked. Then, we only need  to compute the distance from $\boldsymbol{p_c}$ to every point and choose  $N$ out of $L$ with the shortest distance from it. As only one set of distances needs to be calculated, we get linear complexity with $L$ ($\mathcal{O}(L)$). In contrast, an exhaustive search has complexity $\mathcal{O}(L^2)$.

\section{Performance Evaluation}

\subsection{Simulation results}
\begin{figure}[b!]
\centering
  	\includegraphics[width=\linewidth]{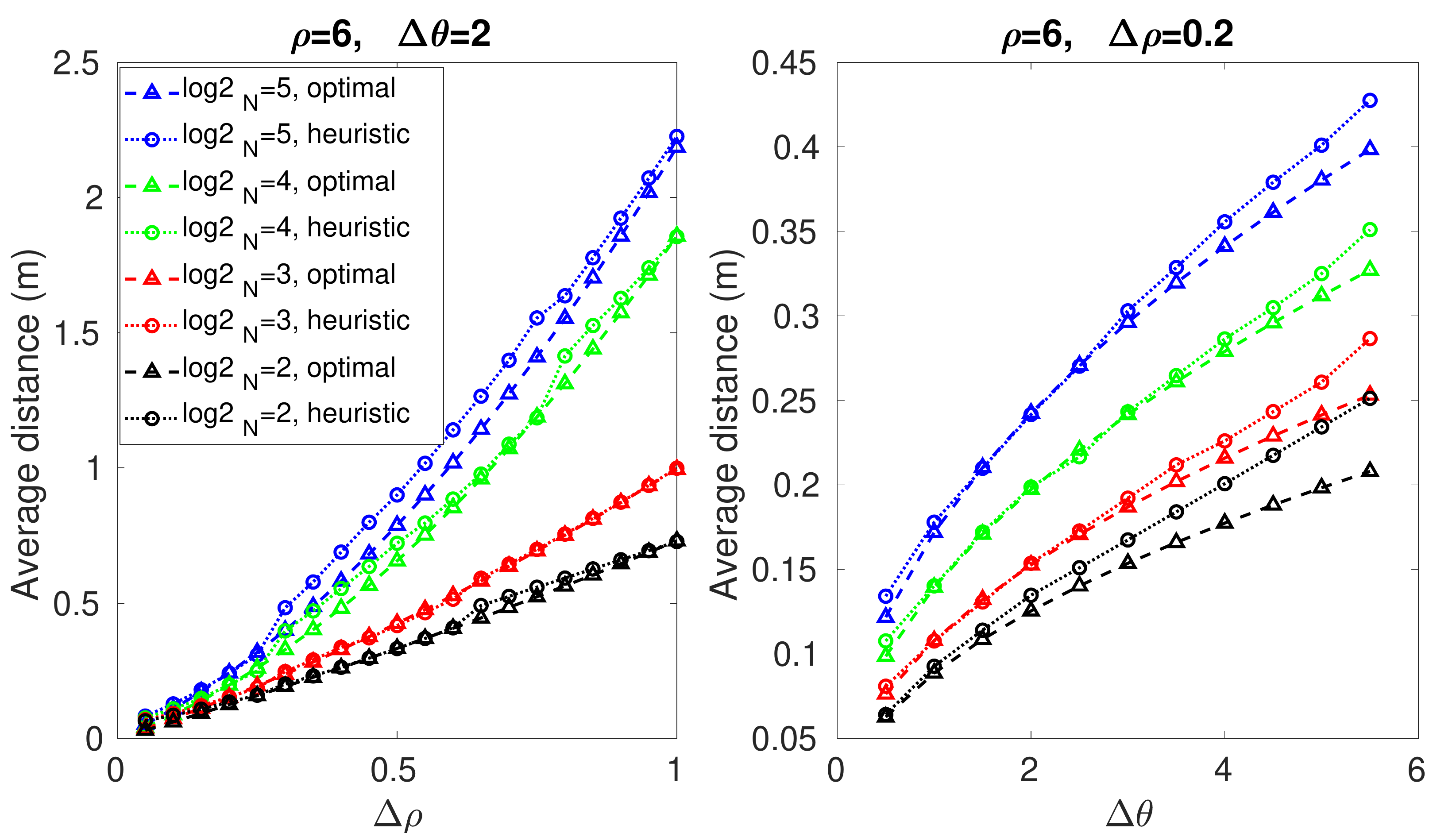}
\caption{Performance of heuristic with respect to optimal }
\label{fig:optimal1}
\end{figure}

\begin{figure}[h!]
    \centering
    \includegraphics[width=\linewidth]{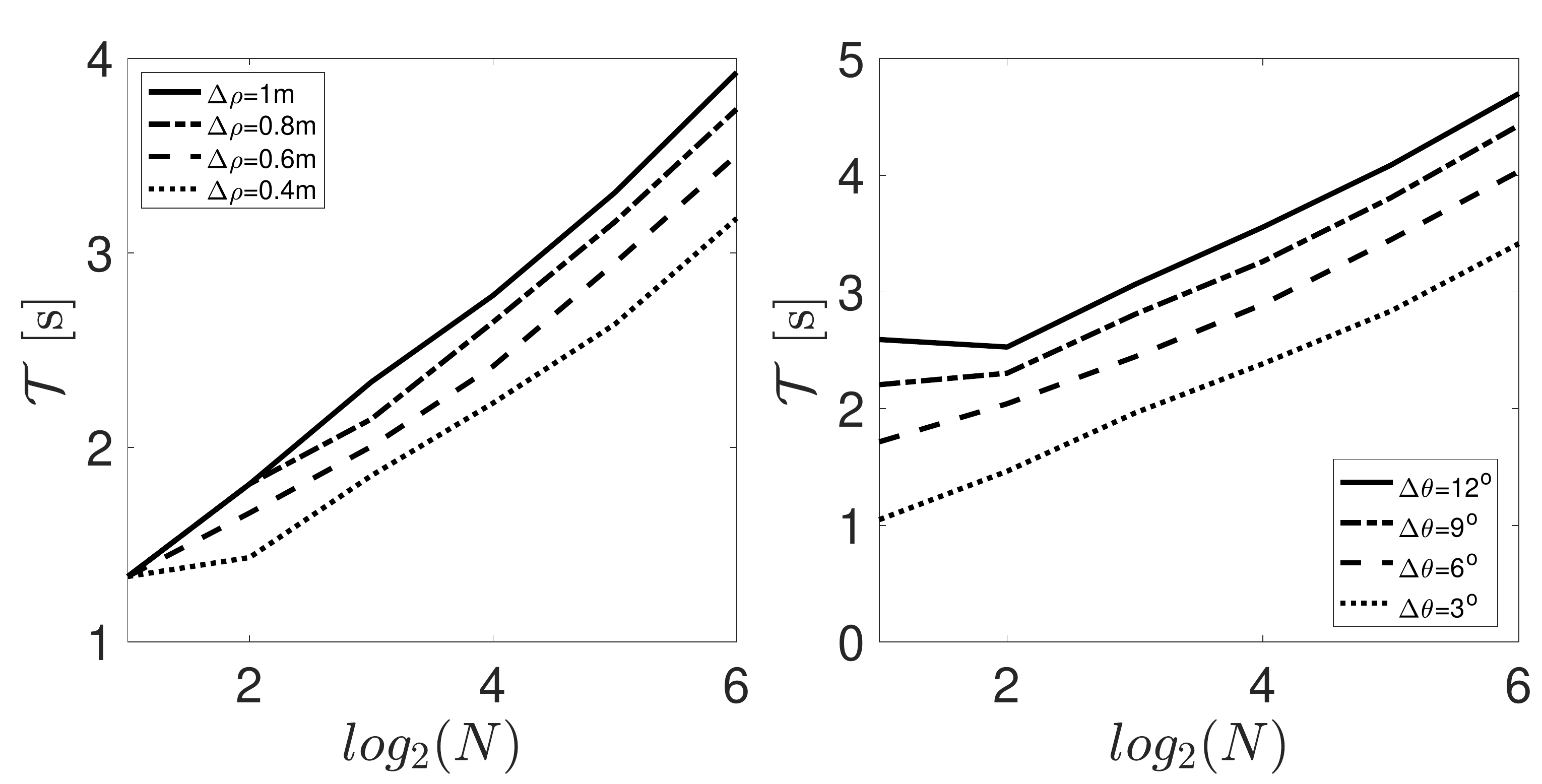}
    \caption{Travel time vs $log_2(N)$ for different $\Delta_{\rho}$ and $\Delta_{\theta}$ values. Both plots are for $\rho = 5m$. On the left, $\Delta_\theta$ is fixed to $5^{o}$. On the right, $\Delta_\rho$ is fixed to 0.8m.}
    \label{fig:times}
    \vspace{-15pt}
\end{figure}

In Fig.~\ref{fig:optimal1}, we compare the average distance for the optimal and heuristically derived constellation points, and we see a near-perfect agreement comparing both approaches. Fig.~\ref{fig:times} analyzes the impact on $\mathcal{T}$ of $\Delta_\theta$ and $\Delta_\rho$, as well as the constellation size. Whenever $\Delta_\theta$ or $\Delta_\rho$ increases, so does $\mathcal{T}$. This is expected since the UAV takes more time for traversing longer inter-symbol distances. The travel time is computed considering a PID controller that chooses the velocity of the UAV based on the distance to its target. We set $k_{p} =0.6$, $k_{d} = 0.12$ and $ k_{i} = 0.05$, which represent for the  \textit{proportional}, \textit{derivative} and \textit{integral} PID constants, respectively. All the simulations are conducted in MATLAB.

\subsection{Experimental results}

\begin{figure}[t!]
    \centering
    \includegraphics[width=0.8\linewidth]{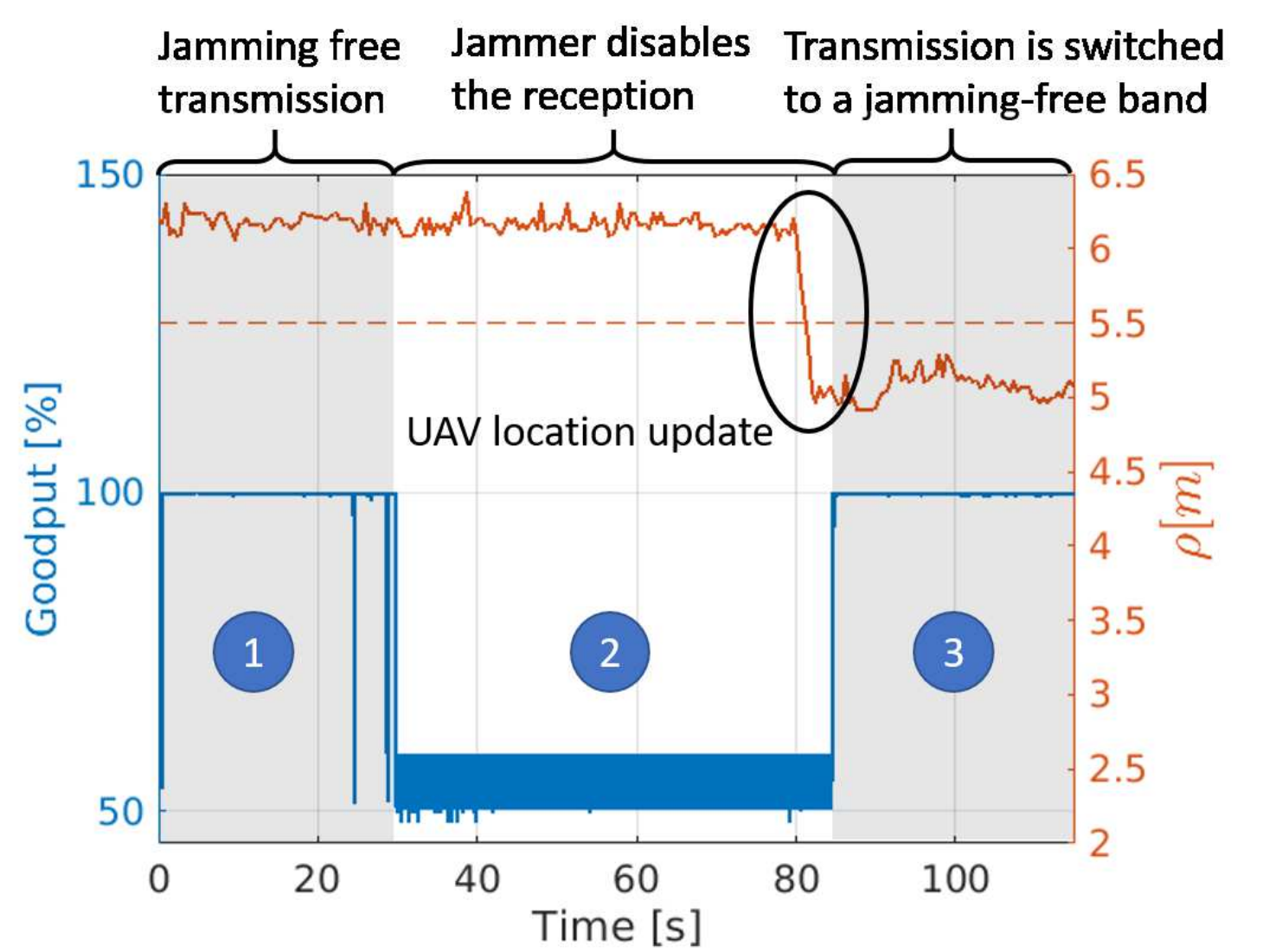}
    \caption{UAV and BS communicate on a certain channel (1) which is jammed right after (2). Then, the UAV moves to a new location within its constellation (from $\rho=6$ to $\rho=5$) to inform the BS the communication will switch to a new band. In (3), the transmission resumes in the new band.}
    \label{fig:BerJamming}
    \vspace{-15pt}
\end{figure}

We assume the problem involves selecting one of two channels centered at 900MHz and 905MHz, and so, we set N=2. For this experiment, we set $\Delta_\rho$ and $\Delta_\theta$ such that $P_e\approx0$. This is achieved by choosing $\Delta_{\rho/ \theta} \gg \sigma_{\rho/ \theta}$. For measurements  with RTK (Sec.~\ref{subsec:pdfs}) we find  $\sigma_\rho\approx0.05m$ and $\sigma_\theta \approx 0.9^o$. Thus, we pick the quotient $\frac{\Delta_{\rho/ \theta}}{\sigma_{\rho/ \theta}}$ to be $\approx$ 13dB, which results in $\Delta_\rho=1m$ and $\Delta_\theta=18^o$. The optimal constellation for these values results in two symbols along the $\rho$ axis. For $\rho=5m$, these points are at $s_1=[0, 5]$ and $s_2=[0, 6]$, where $s_i=[\theta_i, \rho_i]$ represent the coordinates for symbol $i$. Then, we mount one Ettus B210 on a DJI M600 that acts as a transmitter. 
Another B210 and a TI IWR1642 radar is connected to a BS running Linux, which also executes the clustering algorithm from  Sec.~\ref{subsec:Hi-DBSCAN}. The UAV location is an input for switching the center frequency of the B210 SDR. Finally, a third B210 on the ground emulates a jammer by transmitting at high power. Fig.~\ref{fig:BerJamming} shows how goodput evolves over the jamming attack, and how position-based information relaying allows the link to recover via channel switching. Prior to the jamming (1), the average goodput is 100\% with near-perfect data decoding. However, when the jamming attack begins, the receiver is not able to decode the received data (2). Then, the UAV switches the transmission to a new channel and moves to a new constellation point ($\rho=6 \rightarrow \rho=5$) in order to communicate it to the BS. This movement is sensed and the  data transmitted in the new jamming-free channel is decoded at the BS.

\begin{figure}[t!]
    \centering
    \includegraphics[width=0.7\linewidth]{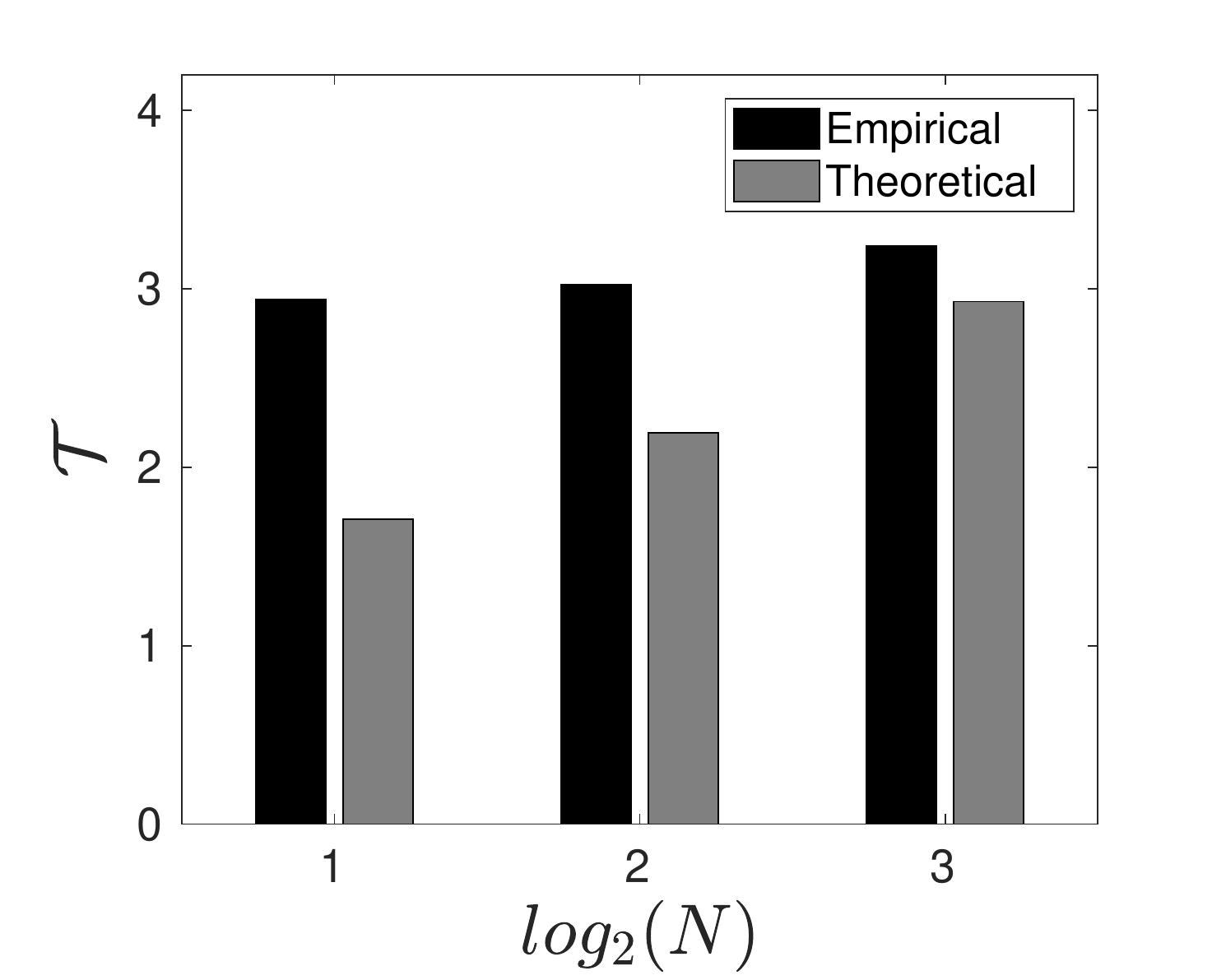}
    \caption{Average travel time comparison. }
    \label{fig:TimevsM}
    \vspace{-20pt}
\end{figure}

Finally, in Fig.~\ref{fig:TimevsM} we show the experimental and theoretical average travel time $\mathcal{T}$. 
We see that our model is more accurate for larger ($\ge N$) constellation sizes. This is because the theoretical model does not account for wind, which makes the UAV take longer time to converge to its next location. Larger constellation sizes are analyzed via simulations in the previous subsection. (Fig.~\ref{fig:times}).

\section{Conclusion}
In this work, we present an interdisciplinary paradigm of position based modulation using UAVs which can convey information by creating spacial codes. As a use case, we show this method can be used for channel selection in jamming situations. Our work is driven by experimental characterization of localization error caused by hovering UAVs, and errors introduced by  COTS mmWave sensor. Finally, we experimentally demonstrate how our system can be used to overcome jamming using a DJI M600, Ettus B210 SDRs and a TI IWR1642 mmWave sensor.

\section*{Acknowledgments}
\vspace{-3pt}
This work is supported by the Office of Naval Research under grant N000141612651.

\bibliographystyle{IEEEtran}
\bibliography{sigproc} 

\begin{thebibliography}{10}
\providecommand{\url}[1]{#1}
\csname url@samestyle\endcsname
\providecommand{\newblock}{\relax}
\providecommand{\bibinfo}[2]{#2}
\providecommand{\BIBentrySTDinterwordspacing}{\spaceskip=0pt\relax}
\providecommand{\BIBentryALTinterwordstretchfactor}{4}
\providecommand{\BIBentryALTinterwordspacing}{\spaceskip=\fontdimen2\font plus
\BIBentryALTinterwordstretchfactor\fontdimen3\font minus
  \fontdimen4\font\relax}
\providecommand{\BIBforeignlanguage}[2]{{%
\expandafter\ifx\csname l@#1\endcsname\relax
\typeout{** WARNING: IEEEtran.bst: No hyphenation pattern has been}%
\typeout{** loaded for the language `#1'. Using the pattern for}%
\typeout{** the default language instead.}%
\else
\language=\csname l@#1\endcsname
\fi
#2}}
\providecommand{\BIBdecl}{\relax}
\BIBdecl

\bibitem{Mozaffari2018}
M.~Mozaffari, W.~Saad, M.~Bennis, Y.~Nam, and M.~Debbah, ``A tutorial on {UAV}s
  for wireless networks: Applications, challenges, and open problems,''
  \emph{CoRR}, vol. abs/1803.00680, 2018.

\bibitem{Wang2017}
J.~Wang, C.~Jiang, Z.~Han, Y.~Ren, R.~G. Maunder, and L.~Hanzo, ``Taking drones
  to the next level: Cooperative distributed unmanned-aerial-vehicular networks
  for small and mini drones,'' \emph{IEEE Vehicular Technology Magazine},
  vol.~12, no.~3, pp. 73--82, Sep. 2017.

\bibitem{Mozaffari2016}
M.~Mozaffari, W.~Saad, M.~Bennis, and M.~Debbah, ``Optimal transport theory for
  power-efficient deployment of unmanned aerial vehicles,'' \emph{CoRR}, vol.
  abs/1602.01532, 2016.

\bibitem{Seo2017}
S.~H. Seo, J.~I. Choi, and J.~Song, ``Secure utilization of beacons and {UAV}s
  in emergency response systems for building fire hazard,'' \emph{Sensors
  (Basel)}, vol.~17, no.~10, pp. 1--21, Sep. 2017.

\bibitem{Trotta2018}
A.~{Trotta}, M.~D. {Felice}, F.~{Montori}, K.~R. {Chowdhury}, and L.~{Bononi},
  ``Joint coverage, connectivity, and charging strategies for distributed uav
  networks,'' \emph{IEEE Transactions on Robotics}, vol.~34, no.~4, pp.
  883--900, Aug 2018.

\bibitem{Xu2005}
W.~Xu, W.~Trappe, Y.~Zhang, and T.~Wood, ``The feasibility of launching and
  detecting jamming attacks in wireless networks,'' in \emph{Proceedings of the
  6th ACM international symposium on Mobile ad hoc networking and
  computing}.\hskip 1em plus 0.5em minus 0.4em\relax ACM, 2005, pp. 46--57.

\bibitem{Naderi2014}
M.~Y. {Naderi}, K.~R. {Chowdhury}, S.~{Basagni}, W.~{Heinzelman}, S.~{De}, and
  S.~{Jana}, ``Experimental study of concurrent data and wireless energy
  transfer for sensor networks,'' in \emph{2014 IEEE Global Communications
  Conference}, Dec 2014, pp. 2543--2549.

\bibitem{Reusmuns2019}
G.~R. Muns, K.~V. Mishra, C.~B. Guerra, Y.~C. Eldar, and K.~R. Chowdhury,
  ``Beam alignment and tracking for autonomous vehicular communication using
  ieee 802.11 ad-based radar,'' in \emph{2019 IEEE INFOCOM Workshop}, May 2019.

\bibitem{Ester1996}
M.~Ester, H.-P. Kriegel, J.~Sander, X.~Xu \emph{et~al.}, ``A density-based
  algorithm for discovering clusters in large spatial databases with noise.''
  \emph{Kdd}, vol.~96, no.~34, pp. 226--231, 1996.

\bibitem{Schubert2017}
E.~Schubert, J.~Sander, M.~Ester, H.~P. Kriegel, and X.~Xu, ``Dbscan revisited,
  revisited: why and how you should (still) use dbscan,'' \emph{ACM
  Transactions on Database Systems (TODS)}, vol.~42, no.~3, p.~19, 2017.

\end{thebibliography}

\end{document}